\documentclass[11pt,a4paper]{article}
\usepackage[T1]{fontenc}
\usepackage{lmodern}
\usepackage[margin=27mm]{geometry}
\usepackage{amsmath,amssymb,amsthm,mathtools,microtype}
\usepackage[hidelinks]{hyperref}
\usepackage[capitalise,nameinlink,noabbrev]{cleveref}

\numberwithin{equation}{section}
\newtheorem{theorem}{Theorem}[section]
\newtheorem{lemma}[theorem]{Lemma}
\newtheorem{corollary}[theorem]{Corollary}
\newtheorem{proposition}[theorem]{Proposition}
\theoremstyle{definition}
\newtheorem{definition}[theorem]{Definition}
\newtheorem{properties}[theorem]{Standard properties}
\newtheorem{example}[theorem]{Example}
\newtheorem{remark}[theorem]{Remark}
\crefname{properties}{Properties}{Properties}
\AddToHook{env/lemma/begin}{\crefalias{theorem}{lemma}}
\AddToHook{env/corollary/begin}{\crefalias{theorem}{corollary}}
\AddToHook{env/proposition/begin}{\crefalias{theorem}{proposition}}
\AddToHook{env/definition/begin}{\crefalias{theorem}{definition}}
\AddToHook{env/properties/begin}{\crefalias{theorem}{properties}}
\AddToHook{env/example/begin}{\crefalias{theorem}{example}}
\AddToHook{env/remark/begin}{\crefalias{theorem}{remark}}

\newcommand{\C}{\mathbb C}
\newcommand{\trans}{\mathsf T}
\newcommand{\Res}{\operatorname*{Res}}
\newcommand{\rem}{\operatorname{rem}}
\newcommand{\fr}{\mathfrak r}

\title{Frobenius--Stickelberger-type determinant identities for $(n,s)$ curves with one unrestricted Abelian argument}
\author{Graham Hesketh}
\date{}

\begin{document}
\maketitle

\begin{abstract}
Frobenius--Stickelberger identities express products of elliptic sigma
functions as determinants. We develop a corresponding formula for $(n,s)$
curves in which one argument vector is unrestricted, while the remaining
arguments come from points on the curve. The determinant's top rows contain curve-point
coordinates, and its bottom rows are explicit linear combinations of
Kleinian functions and their derivatives. We give a finite procedure for
constructing these rows and illustrate it with hyperelliptic and trigonal
examples. For Baker--Akhiezer modes on odd-degree hyperelliptic curves,
the determinant can also be written polynomially in elementary mode powers
and their first derivatives, providing a tool for studying wave-mixing
products and conservation laws.
\end{abstract}

\section{Introduction}\label{sec:introduction}

Determinant identities connecting products of sigma functions with elliptic
functions have a long history, including the nineteenth-century formulas
of Kiepert \cite{Kiepert1873}, Hermite \cite{Hermite1877}, and
Frobenius--Stickelberger \cite{FrobeniusStickelberger1877}. \^Onishi extended these formulas to
hyperelliptic curves, with the arguments restricted to Abel lifts of curve
points \cite{Onishi2005}. Nakayashiki subsequently obtained sigma-function
determinant formulas for $(n,s)$ curves \cite{Nakayashiki2008}. Related
addition identities appear in Fay's theta-function theory \cite{Fay1973}
and in determinant and Wronskian formulas for Baker--Akhiezer (BA) functions
\cite{Nakayashiki2026}. These results provide complementary ways to express
products of special functions through algebraic curve data.
Within this algebraic-function approach, Kleinian Jacobi inversion is
developed for canonical curves in \cite{EEL2000}, with explicit genus-three
trigonal equations in \cite{EEMOP2007}; \^Onishi's purely trigonal
determinant likewise uses curve-point arguments
\cite[Theorem~4.3]{Onishi2009Trigonal}.

One motivation for such identities comes from integrable wave interactions.
In the author's work on elliptic optical systems, Frobenius--Stickelberger
identities express the mixing products entering canonical Hamiltonians in
terms of modal powers and their derivatives, giving a functional
explanation of the Hamiltonian and Manley--Rowe-type conservation laws
\cite{Hesketh2015,HeskethFWM2026}. For higher-order mixing products,
\^Onishi's hyperelliptic determinants play a corresponding role when the
arguments remain Abel lifts of curve points \cite{HeskethHyperelliptic2026}.
More general BA dynamics requires an unrestricted argument $z\in\C^g$,
while the elementary shifts remain associated with points of the curve.
This is the setting encountered in the canonical three-wave systems
studied in \cite{HeskethBA}, and is also relevant to the moment-map and KP
constructions discussed there.

We therefore seek a Frobenius--Stickelberger-type identity with one
unrestricted Abelian argument. For a smooth $(n,s)$ curve of genus $g$, we
express a sigma product involving $z$, elementary shifts and their total
shift as a bordered determinant. Its first rows contain only the fixed
curve-point coordinates, in the manner of \^Onishi's presentation; its
bottom $g$ rows contain explicit linear combinations of Kleinian $\wp$
functions and their derivatives at $z$, together with curve-dependent
constants. The determinant evaluation follows from established addition
formulas. Our contribution is a uniform generating-function formula for
every remainder column, combined with this separated determinant and its
point-derivative and mode-power forms. Nakayashiki's degenerate addition
formulas \cite[Theorem~6.2]{Nakayashiki2026} instead present the shifted
function through BA/Wronskian entries; here the bottom rows are explicit
linear Kleinian expressions at $z$.

The resulting identities express products of elementary BA modes and a
total-shift composite through Abelian functions of the evolving argument.
They provide tools for rewriting mixing terms and studying polynomial
conservation laws from the function perspective. A complementary
point-derivative form makes the connection with the derivative columns of
the classical elliptic formulas explicit.

\Cref{sec:setting} fixes the curve and function conventions.
\Cref{sec:bordered} establishes the bordered determinant evaluation.
\Cref{sec:main} derives the Kleinian row formula and its point-derivative
form. \Cref{sec:mode-powers} gives the hyperelliptic mode-power
representation, and \cref{sec:examples} presents genus-two and trigonal
examples.

\section{Setting}\label{sec:setting}

\begin{definition}[Curve, monomials and differentials]\label{def:curve}
For coprime integers $n,s\geq2$, an \emph{$(n,s)$ curve} $X$ is the
compactification, by one point $\infty$, of a smooth affine curve
\[
F(x,y)=y^n-c\,x^s-\sum_{an+bs<ns}\lambda_{ab}\,x^ay^b=0,\qquad
\gcd(n,s)=1,\quad c\ne0,
\]
of genus $g=\tfrac12(n-1)(s-1)$, where the sum runs over nonnegative
integers $a,b$ satisfying $an+bs<ns$; the constant $c$ allows conventions such as
$y^2=4x^5+\cdots$. Fix one of the $n$ choices of local parameter $t$ at $\infty$ with
$x=t^{-n}$; this fixes the leading coefficient $\ell_y$ of $y$ and the
local normalization constants.
The monomials $x^ay^b$, $0\le b\le n-1$, ordered by their pole orders
$w_1<w_2<\cdots$ at $\infty$, are denoted $\varphi_1=1,\varphi_2,\varphi_3,\dots$,
and $\ell_k$ is the leading coefficient of $\varphi_k$:
$\varphi_k=\ell_k\,t^{-w_k}(1+O(t))$. Put $\nu=c_\nu\,dx/F_y$ for a fixed
constant $c_\nu\ne0$ ($c_\nu=1$ unless stated otherwise), a differential
with divisor $(2g-2)\infty$. Then
\[
du_j=\varphi_j\,\nu,\qquad j=1,\dots,g,
\]
are holomorphic differentials, forming a basis, and for every polynomial
$\varphi$ in $x,y$ the differential $\varphi\,\nu$ has no pole except at
$\infty$. This ascending ordering is the Baker basis of \cite{BEL1997}, used
in our earlier work \cite{HeskethBA} and in the numerical code \cite{Generapy};
for $y^2=4x^5+\cdots$ and $c_\nu=2$ it gives $du_1=dx/y$, $du_2=x\,dx/y$.
Nakayashiki \cite[p.~184]{Nakayashiki2008} orders the basis in reverse and with
the opposite sign, $-\varphi_{g+1-i}\,dx/F_y$. All statements below are in our
conventions.
\end{definition}

\begin{definition}[Sigma and Kleinian functions]\label{def:sigma}
Let $dr_1,\dots,dr_g$ be second-kind differentials with poles only at
$\infty$, normalized by
\begin{equation}\label{eq:legendre}
\Res_\infty(u_i\,dr_j)=\delta_{ij},\qquad
\Res_\infty(\fr_i\,dr_j)=0,\qquad 1\le i,j\le g.
\end{equation}
Here $u_i$ and $\fr_i$ are local primitives of $du_i$ and $dr_i$ near
$\infty$. These conditions permit symmetric holomorphic shifts
$dr\mapsto dr+Hdu$, $H=H^\trans$. We fix one choice throughout:
the Baker/BEL choice for hyperelliptic curves, as specified in this definition,
and the stated choices in the other examples. For general $(n,s)$ curves
any fixed choice satisfying \eqref{eq:legendre} may be used. Let $2\omega,2\omega'$ be the period matrices of $du$ and
$-2\eta,-2\eta'$ those of $dr$, and $\varkappa=\eta\omega^{-1}$, which is
symmetric by \eqref{eq:legendre}. The \emph{sigma function} is
\[
\sigma(u)=C\,e^{\frac12u^\trans\varkappa u}\,
\theta[\delta]\bigl((2\omega)^{-1}u\,\big|\,\omega^{-1}\omega'\bigr),\qquad u\in\C^g,
\]
with $\delta$ the characteristic of the vector of Riemann constants
with base point $\infty$. The constant
$C$ fixes the multiplicative normalization, separately from the choice of $dr$. To record its relation to the
Schur normalization of \cite{NakayashikiYori2012}, proceed as follows.
Let $\beta_j=2g-1-w_j$, the vanishing order of $u_j$ at $\infty$, so
that $\beta_1>\cdots>\beta_g=1$ are the gaps at $\infty$; let $\lambda$ be the
partition with $\lambda_j=\beta_j-g+j$; and put $\varepsilon=c_\nu\,\ell_y^{1-n}$,
where $y=\ell_y\,t^{-s}(1+O(t))$. Then
$u_j=-\varepsilon\ell_j\,t^{\beta_j}/\beta_j\,(1+O(t))$ near $\infty$. Define
$T(u)=(T_1,T_2,\dots)$ by $T_{\beta_j}=-u_j/(\varepsilon\ell_j)$ and $T_m=0$ for
$m$ not a gap. Then
\[
\begin{gathered}
\sigma(u)=c_\sigma s_\lambda\bigl(T(u)\bigr)+\cdots,\qquad
s_\lambda(T)=\det\bigl(p_{\lambda_i-i+j}(T)\bigr)_{i,j=1}^g,\\
\sum_{m\ge0}p_m(T)\,\chi^m=\exp\Bigl(\sum_{k\ge1}T_k\chi^k\Bigr),
\end{gathered}
\]
where $p_m$ are the one-row Schur polynomials defined by the generating
function, with $p_m=0$ for $m<0$, and $s_\lambda$ is the Schur polynomial
of partition $\lambda$. Here $\chi$ is a formal variable, $c_\sigma\ne0$, and the omitted terms
have higher weighted degree in $u$. We use Sato weights
$\operatorname{wt}(u_j)=\beta_j$, $\operatorname{wt}(x)=-n$,
$\operatorname{wt}(y)=-s$ and
$\operatorname{wt}(\lambda_{ab})=an+bs-ns$; $c,c_\nu$ have weight zero. On hyperelliptic curves
$y^2=4x^{2g+1}+\cdots$, with $du_j=x^{j-1}dx/y$, we take the Baker/BEL polynomial fundamental
form and its associated second-kind differentials \cite{BEL1997,EEL2000}.
The multiplicative Baker normalization fixes the lowest ordinary-degree
term of $\sigma$ to be
\[
\det[u_{i+j-1}]_{i,j=1}^{\lfloor(g+1)/2\rfloor}.
\]
This Baker convention \cite{BEL1997} agrees with \cite{HeskethBA} and the hyperelliptic
normalization implemented by Generapy, the open-source Python library
we use for numerical validation \cite{Generapy}. It fixes
$c_\sigma$ relative to the chosen local parameter. For example, with
$y\sim2t^{-(2g+1)}$, $c_\sigma=-1$ in genus one and $c_\sigma=1$ in genus
two: $\sigma(u)=u+\cdots$ and
$\sigma(u)=u_1-\tfrac13u_2^3+\cdots$, respectively. For other $(n,s)$
curves we take $c_\sigma=1$ unless stated otherwise. For the same period
data, Generapy's default non-hyperelliptic theta scaling is another
multiplicative normalization;
the balanced identity \eqref{eq:det} is unchanged by such a rescaling.
The \emph{Kleinian functions} are
\[
\zeta_j=\partial_j\log\sigma,\qquad
\wp_{j_1\cdots j_r}=-\partial_{j_1}\cdots\partial_{j_r}\log\sigma\quad(r\ge2),
\]
with $\partial_j=\partial/\partial u_j$ and $\partial=(\partial_1,\dots,\partial_g)$.
\end{definition}

\begin{definition}[Abel map and remainders]\label{def:remainder}
Let $A(P)=\int_\infty^Pdu\in\C^g$, and let $\fr_j$ be a primitive of $dr_j$,
continued along the same paths as $A$; its additive constant plays no role.
For a generic effective divisor $D$ of degree $g$, its \emph{argument} is
$z=A(D)$. For a polynomial $\varphi$ in $x,y$, the \emph{remainder}
$\rem_z(\varphi)\in\C^g$ is the vector for which
\[
\varphi-\sum_{j=1}^g\rem_z(\varphi)_j\,\varphi_j\quad\text{vanishes on }D;
\]
equivalently, $\varphi\,\nu-\sum_j\rem_z(\varphi)_j\,du_j$ vanishes on $D$.
\end{definition}

\begin{definition}[Prime functions]\label{def:prime}
$\tilde E(P,Q)$ and $\tilde E(\infty,P)$ denote the prime functions of
\cite[\S5.3]{Nakayashiki2008} and \cite[\S5]{NakayashikiYori2012}, normalized at $\infty$ with the local
parameter $t$. Their defining quadratic factor uses the same
second-kind period data $\eta,\eta'$ as $\sigma$ in \cref{def:sigma}.
\end{definition}

\begin{properties}\label{prop:standard}
The following are used in the proofs.
\begin{enumerate}
\item \emph{Quasi-periodicity.} For $m,m'\in\mathbb Z^g$,
$\sigma(u+2\omega m+2\omega'm')=\pm\,e^{2(\eta m+\eta'm')^\trans(u+\omega m+\omega'm')}\sigma(u)$,
so $\zeta_j(u+2\omega m+2\omega'm')=\zeta_j(u)+2(\eta m+\eta'm')_j$.
\item \emph{Riemann vanishing} \cite{NakayashikiYori2012}. For a nonspecial effective divisor $D$ of degree $g$, with
argument $z$ and $\sigma(z)\ne0$, the function $P\mapsto\sigma(z-A(P))$ vanishes exactly on $D$.
\item \emph{Prime functions} \cite[Proposition~7, Proposition~8, Lemma~16]{Nakayashiki2008}. As a function of
$P$, $\tilde E(P,Q)$ has zeros exactly at $Q+(g-1)\infty$, and
$\tilde E(Q,P)=-\tilde E(P,Q)$; $\tilde E(\infty,P)$ has zeros exactly at
$g\infty$. $\tilde E(P,Q)$ and $\tilde E(\infty,P)$ have the multipliers of
$\sigma(A(Q)-A(P))$ and $\sigma(A(P))$. Near $\infty$,
$\tilde E(\infty,P)=t^g+O(t^{g+1})$ and
$\tilde E(P,Q)=-\tilde E(\infty,Q)\,t(P)^{g-1}+O(t(P)^g)$.
Both are derivatives of $\sigma$ \cite[Theorem~10, Corollary~6]{NakayashikiYori2012}:
with $N=\lambda_2+\cdots+\lambda_g$ and $N'=N-g+1$, $\tilde E(\infty,P)$ is a
constant multiple of $\partial_g^{N}\sigma(A(P))$ and $\tilde E(P,Q)$ one of
$\partial_g^{N'}\sigma(A(P)-A(Q))$. \Cref{cor:closed} gives the constants.
\item \emph{Nakayashiki's determinant formula}
\cite[Theorem~1]{Nakayashiki2008}; see also \cite[Theorem~11]{NakayashikiYori2012}.
For $q\ge g$ points $p_1,\dots,p_q$,
\[
\det\bigl[\varphi_c(p_r)\bigr]_{r,c=1}^q=\frac{1}{c_\sigma}\prod_{c=1}^q\ell_c\cdot
\sigma\Bigl(\sum_{r=1}^qA(p_r)\Bigr)\,
\frac{\prod_{1\le r<r'\le q}\tilde E(p_r,p_{r'})}{\prod_{r=1}^q\tilde E(\infty,p_r)^q},
\]
where $c_\sigma^{-1}$ converts the normalization of \cref{def:sigma}
to that of the cited formula, and the factors $\ell_c$ account for
non-monic leading terms. For
hyperelliptic curves, \^Onishi's generalized determinant formula
\cite[Theorem~7.2]{Onishi2005} is a formula of the same kind.
\end{enumerate}
\end{properties}

\section{The bordered determinant}\label{sec:bordered}

We evaluate the determinant with abstract remainder rows using Nakayashiki's
formula and a one-variable divisor comparison. The auxiliary product
identity is classical \cite[Lemma~2.7, (34)]{Fay1973}; the proof below
establishes it directly and fixes its constant.

For distinct finite curve points $P_1,\dots,P_K$, with $\mu_i=A(P_i)$,
and a generic nonspecial degree-$g$ divisor $D=Q_1+\cdots+Q_g$, whose
points $Q_l$ lie on $X$ and whose Abel argument is $z$, put
\[
\mathsf V=[\varphi_j(P_i)]_{\substack{1\le i\le K\\1\le j\le g}},\qquad
\mathsf W=[\varphi_{g+k}(P_i)]_{1\le i,k\le K},\qquad
\mathsf M_z=[\rem_z(\varphi_{g+k})_j]_{\substack{1\le j\le g\\1\le k\le K}}.
\]
\begin{lemma}[Bordered evaluation]\label{lem:bordered}
\begin{equation}\label{eq:bordered}
\begin{aligned}
(-1)^{gK}\det\begin{pmatrix}\mathsf V&\mathsf W\\\mathsf I_g&\mathsf M_z\end{pmatrix}
&=\det(\mathsf W-\mathsf V\mathsf M_z)\\
&=\Bigl(\prod_{k=g+1}^{g+K}\ell_k\Bigr)
\frac{\sigma(z+\sum_{i=1}^K\mu_i)\prod_{i=1}^K\sigma(z-\mu_i)}{\sigma(z)^{K+1}}
\frac{\prod_{1\le i<j\le K}\tilde E(P_i,P_j)}{\prod_{i=1}^K\tilde E(\infty,P_i)^{K+1}}.
\end{aligned}
\end{equation}
This is an identity of meromorphic functions of $z$.
\end{lemma}
\begin{proof}
Write $D=Q_1+\cdots+Q_g$, $e(P)=\tilde E(\infty,P)$ and
$B=[\varphi_j(Q_l)]_{l,j=1}^g$. Multiplying the bottom block by $B$
replaces it by the evaluation rows at $D$, giving
\[
\det\begin{pmatrix}\mathsf V&\mathsf W\\\mathsf I_g&\mathsf M_z\end{pmatrix}
=\frac{\det[\varphi_c(p_r)]_{r,c=1}^{K+g}}
       {\det[\varphi_j(Q_l)]_{l,j=1}^g},\qquad
(p_1,\dots,p_{K+g})=(P_1,\dots,P_K,Q_1,\dots,Q_g).
\]
To evaluate the cross factors, fix $D$ and vary only $R$. Continue all
Abel integrals and prime functions along the same paths. Under a period
$\Lambda$, write $\mathsf m_\Lambda(u)=\mathrm{sgn}_\Lambda
 e^{2v_\Lambda^\trans(u+\Lambda/2)}$, with
$\mathrm{sgn}_{-\Lambda}=\mathrm{sgn}_\Lambda\in\{1,-1\}$.
For $r=A(R)$ both $\prod_l\tilde E(Q_l,R)$ and
$e(R)^{g-1}\prod_l e(Q_l)\sigma(z-r)/\sigma(z)$ acquire the multiplier
\[
 \mathrm{sgn}_\Lambda^g
 e^{2v_\Lambda^\trans(gr-z+g\Lambda/2)}.
\]
Antisymmetry and \cref{prop:standard}(3) give the zero divisor
$D+g(g-1)\infty$ for the product in its second argument $R$; the other
expression has the same divisor. Their quotient
is therefore constant, and the local expansions at $R=\infty$ give
constant one. Consequently
\[
 \prod_{l=1}^g\tilde E(Q_l,R)
 =e(R)^{g-1}\prod_{l=1}^g e(Q_l)\frac{\sigma(z-A(R))}{\sigma(z)}.
\]
Nakayashiki's formula in \cref{prop:standard} gives
\[
\begin{aligned}
\frac{\det[\varphi_c(p_r)]_{r,c=1}^{K+g}}
     {\det[\varphi_j(Q_l)]_{l,j=1}^g}
&=\Bigl(\prod_{k=g+1}^{g+K}\ell_k\Bigr)
\frac{\sigma(z+\sum_i\mu_i)}{\sigma(z)}\\
&\quad\times
\frac{\prod_{i<j}\tilde E(P_i,P_j)\prod_{i,l}\tilde E(P_i,Q_l)}
     {\prod_i e(P_i)^{K+g}\prod_l e(Q_l)^K}.
\end{aligned}
\]
The divisor-only factors and $c_\sigma$ have canceled. Substituting the
auxiliary identity with $R=P_i$ leaves powers $-(K+1)$ of $e(P_i)$.
Reversing the $gK$ cross prime functions gives $(-1)^{gK}$, proving the
last equality. Subtracting the first $g$ columns times $\mathsf M_z$ from
the last $K$ gives the Schur-complement equality. Meromorphic continuation
extends the identity from the generic chart.
\end{proof}

The main theorem now computes the remainder block explicitly in Kleinian
functions, without requiring the points of its inversion divisor.

\section{Main theorem}\label{sec:main}

For a polynomial $\varphi$ in $x,y$, write the principal part of
$\varphi\nu$ at $\infty$ as
\begin{equation}\label{eq:principal}
\varphi\,\nu=\sum_{m=1}^{d-1}c_m(\varphi)\,t^{-m-1}dt+O(1),
\end{equation}
where $d$ is the pole order of $\varphi\nu$ (the residue vanishes, since
$\infty$ is the only pole). If it is holomorphic, set $d=1$ and read the
principal-part sum as empty. Define operators $\mathcal D_m(\partial)$ by
\begin{equation}\label{eq:generating}
\exp\bigl(-A(P)\cdot\partial\bigr)\Big|_{t(P)=\chi}=\sum_{m=0}^{\infty}\mathcal D_m(\partial)\,\chi^m .
\end{equation}
Explicitly, if $A(P)=\sum_{k\ge1}\alpha_k\,t(P)^k$ with $\alpha_k\in\C^g$, then
$\mathcal D_m(\partial)=p_m(-\alpha_1\cdot\partial,-\alpha_2\cdot\partial,\dots)$, with
the Schur polynomials $p_m$ of \cref{def:sigma}. Here $\chi$ is a formal variable. For $m\ge1$ we write
\[
\mathcal D_m(\partial)=\sum_{r=1}^m\sum_{i_1,\dots,i_r=1}^g\pi_{m;i_1\cdots i_r}\,\partial_{i_1}\cdots\partial_{i_r}.
\]
Here $\pi_{m;i_1\cdots i_r}$ are the constant coefficients in this
differential-operator expansion, determined by the local Abel series.

\begin{theorem}[Determinant formula with linear Kleinian rows]\label{thm:main}
\leavevmode
\begin{enumerate}
\item For every polynomial $\varphi$ in $x,y$, the argument $z$ of a generic
divisor, and $j=1,\dots,g$,
\begin{equation}\label{eq:rows}
\rem_z(\varphi)_j
=\sum_{m=1}^{d-1}c_m(\varphi)\,\partial_j\mathcal D_m(\partial)\log\sigma(z)
\;-\;\Res_\infty\bigl(\fr_j\,\varphi\,\nu\bigr),
\end{equation}
that is,
\begin{equation}\label{eq:rows-kleinian}
\rem_z(\varphi)_j
=-\sum_{m=1}^{d-1}c_m(\varphi)\sum_{r=1}^{m}\sum_{i_1,\dots,i_r=1}^g
\pi_{m;i_1\cdots i_r}\,\wp_{j\,i_1\cdots i_r}(z)
-\Res_\infty\bigl(\fr_j\,\varphi\,\nu\bigr).
\end{equation}
\item Consequently, for $K\ge1$ distinct finite generic curve points $P_1,\dots,P_K$ with
$\mu_i=A(P_i)$, and generic $z\in\C^g$,
\begin{equation}\label{eq:det}
\begin{aligned}
&(-1)^{gK}\det\begin{pmatrix}
\bigl[\varphi_j(P_i)\bigr]&\bigl[\varphi_{g+k}(P_i)\bigr]\\[2pt]
\mathsf I_g&\bigl[\mathcal R_j(\varphi_{g+k})(z)\bigr]
\end{pmatrix}\\
&\qquad=\Bigl(\prod_{k=g+1}^{g+K}\ell_k\Bigr)\,
\frac{\sigma\bigl(z+\sum_{i=1}^K\mu_i\bigr)\prod_{i=1}^K\sigma(z-\mu_i)}{\sigma(z)^{K+1}}\,
\frac{\prod_{1\le i<j\le K}\tilde E(P_i,P_j)}{\prod_{i=1}^K\tilde E(\infty,P_i)^{K+1}},
\end{aligned}
\end{equation}
where $1\le i,k\le K$, $1\le j\le g$, and $\mathcal R_j(\varphi)(z)$ is the
right-hand side of \eqref{eq:rows-kleinian}: an explicit linear combination of
Kleinian functions at $z$ plus a constant.
\end{enumerate}
\end{theorem}

In \eqref{eq:det} the first $K$ rows involve only the curve points and the
last $g$ rows only the free argument $z$, linearly through Kleinian functions.
The new ingredient is part~(i); part~(ii) follows from it and
\cref{lem:bordered}. For the holomorphic columns $\varphi_i$,
$1\leq i\leq g$, the empty sum gives
$\mathcal R_j(\varphi_i)=-\Res_\infty(\fr_jdu_i)=\delta_{ij}$,
recovering the identity block.

\begin{proof}
The rows are the Taylor coefficients at $\infty$ of the log-derivative
$\zeta_j(z-A(P))=\partial_{z_j}\log\sigma(z-A(P))$, whose poles in $P$ are the
points of $D$; the residue theorem converts them into remainders. The
mechanism is that of Fay \cite[Lemma~2.7, (32), (35)]{Fay1973}.

\emph{(i)} Let $D=Q_1+\dots+Q_g$ be generic with argument $z$: the $Q_l$ are
distinct, differ from $\infty$, and the matrix $[du_j(Q_l)]$ is invertible.

\emph{Step 1: a single-valued kernel.} When $P$ traverses a cycle, $A(P)$
changes by a period $2\omega m+2\omega'm'$ and $\fr_j(P)$ changes by
$\oint dr_j=-2(\eta m+\eta'm')_j$, while by quasi-periodicity
$\zeta_j(z-A(P))$ also changes by $-2(\eta m+\eta'm')_j$. Hence
\[
\mathcal K_j(P)=\zeta_j\bigl(z-A(P)\bigr)-\fr_j(P)
\]
is a single-valued meromorphic function on $X$. By Riemann vanishing its
poles are the points of $D$ and the point $\infty$. A change of the paths
changes $A$ and $\fr_j$ together and leaves $\mathcal K_j$ unchanged; a change of
the additive constant of $\fr_j$ adds a constant to $\mathcal K_j$, which
multiplies $\Res_\infty(\varphi\nu)=0$ in Step~3.

\emph{Step 2: residues on $D$.} Near $Q_l$, in a local parameter $\tau$,
$\sigma(z-A(P))$ has a simple zero at $\tau=\tau_l(z)$, so
$\zeta_j(z-A(P))=-(\partial\tau_l/\partial z_j)/(\tau-\tau_l)+O(1)$ and
\[
\Res_{Q_l}\bigl(\mathcal K_j\,\varphi\nu\bigr)=-\frac{\partial\tau_l}{\partial z_j}\,\frac{\varphi\nu}{d\tau}(Q_l).
\]
Since $z=\sum_lA(Q_l)$, differentiation gives
$\sum_l\frac{du_i}{d\tau}(Q_l)\,\frac{\partial\tau_l}{\partial z_j}=\delta_{ij}$. On
the other hand $\rem_z(\varphi)$ solves
$\sum_i\rem_z(\varphi)_i\,\frac{du_i}{d\tau}(Q_l)=\frac{\varphi\nu}{d\tau}(Q_l)$
for all $l$. Combining the two,
\[
\sum_{l=1}^g\Res_{Q_l}\bigl(\mathcal K_j\,\varphi\nu\bigr)
=-\sum_{l=1}^g\frac{\partial\tau_l}{\partial z_j}\sum_{i=1}^g\rem_z(\varphi)_i\frac{du_i}{d\tau}(Q_l)
=-\rem_z(\varphi)_j .
\]

\emph{Step 3: the residue theorem.} The residues of $\mathcal K_j\,\varphi\nu$
sum to zero. Its only pole besides $D$ is $\infty$, where
$\zeta_j(z-A(P))$ is regular for generic $z$. Therefore
\[
\rem_z(\varphi)_j=\Res_\infty\bigl(\zeta_j(z-A(P))\,\varphi\nu\bigr)
-\Res_\infty\bigl(\fr_j\,\varphi\nu\bigr).
\]

\emph{Step 4: the generating function.} Near $\infty$, with $t(P)=\chi$ and
$A(\infty)=0$, \eqref{eq:generating} gives
\[
\begin{aligned}
\zeta_j(z-A(P))
&=\exp\bigl(-A(P)\cdot\partial\bigr)\zeta_j(z)\\
&=\sum_{m\ge0}\chi^m\,\partial_j\mathcal D_m(\partial)\log\sigma(z).
\end{aligned}
\]
Pairing with the principal part \eqref{eq:principal} gives
\[
\Res_\infty\bigl(\zeta_j(z-A(P))\,\varphi\nu\bigr)
=\sum_{m=1}^{d-1}c_m(\varphi)\,\partial_j\mathcal D_m(\partial)\log\sigma(z),
\]
which is \eqref{eq:rows}. Since $\wp=-\partial\partial\log\sigma$ and $\mathcal D_m$
has no constant term for $m\ge1$, \eqref{eq:rows} is \eqref{eq:rows-kleinian}.

\emph{(ii)} A generic $z\in\C^g$ is the argument of a generic divisor $D$ (Jacobi
inversion). By part~(i), $\mathsf M_z=[\mathcal R_j(\varphi_{g+k})(z)]$, so
\eqref{eq:det} is \eqref{eq:bordered}.
\end{proof}

The prime functions in \eqref{eq:det} can be written with $\sigma$ alone.
With $N$ and $N'$ as in \cref{prop:standard}, put
\[
\Delta_\beta=\prod_{2\le i<j\le g}(\beta_i-\beta_j),
\]
\[
a_\infty=c_\sigma(-\varepsilon\ell_g)^{-N}\,\frac{N!\,\Delta_\beta}{\prod_{j=2}^g\beta_j!},\qquad
a_{\mathrm{pair}}=c_\sigma(-1)^{g-1}(-\varepsilon\ell_g)^{-N'}\,\frac{N'!\,\Delta_\beta}{\prod_{j=2}^g(\beta_j-1)!}.
\]

\begin{corollary}[Closed form]\label{cor:closed}
$\tilde E(\infty,P)=\partial_g^N\sigma(A(P))/a_\infty$ and
$\tilde E(P,Q)=\partial_g^{N'}\sigma(A(P)-A(Q))/a_{\mathrm{pair}}$. Hence, in \cref{thm:main}(ii),
\[
\begin{aligned}
&(-1)^{gK}\det(\cdots)\\
&\qquad=\Bigl(\prod_{k=g+1}^{g+K}\ell_k\Bigr)\,
\frac{a_\infty^{K(K+1)}}{a_{\mathrm{pair}}^{K(K-1)/2}}\,
\frac{\sigma\bigl(z+\sum_{i=1}^K\mu_i\bigr)\prod_{i=1}^K\sigma(z-\mu_i)}{\sigma(z)^{K+1}}\,
\frac{\prod_{1\le i<j\le K}\bigl[\partial_g^{N'}\sigma(\mu_i-\mu_j)\bigr]}{\prod_{i=1}^K\bigl[\partial_g^N\sigma(\mu_i)\bigr]^{K+1}},
\end{aligned}
\]
with the same determinant as in \eqref{eq:det}.
\end{corollary}

\begin{proof}
Write $y=\ell_yY$. Then $F=c\,G$ with $G$ monic in $Y$ and $x$,
$\nu=\varepsilon\,dx/G_Y$, and $\varphi_j=\ell_j\varphi_j^{G}$ for the monomials
$\varphi_j^G$ of $G$. So $T(A(P))$ is Nakayashiki's Abel map of $G=0$
\cite[p.~184]{Nakayashiki2008}, indexed by gaps, and by \cref{def:sigma}
\[
\sigma(u)=c_\sigma e^{q(u)}\sigma_{\rm NY}(T(u)),
\]
where the fixed quadratic form $q$ accounts for a different admissible
second-kind choice. On the curve and two-point difference loci, respectively, all
lower $\partial_g$ derivatives vanish by
\cite[Theorems~8 and~9(i)]{NakayashikiYori2012}. Thus the first nonzero derivatives
and the corresponding prime functions acquire the same quadratic gauge.
By
\cite[Theorem~10, Corollary~6]{NakayashikiYori2012}, $\partial_g^N\sigma(A(P))$ and
$\partial_g^{N'}\sigma(A(P)-A(Q))$ are constant multiples of $\tilde E(\infty,P)$ and
$\tilde E(P,Q)$. The constants are fixed by the leading terms at $\infty$,
which come from $c_\sigma s_\lambda$. Only gap variables are retained:
\[
\begin{aligned}
T_{\beta_j}(A(P))&=\frac{t^{\beta_j}}{\beta_j}+\cdots,\\
T_{\beta_j}(A(P)-A(Q))&=\frac{t_P^{\beta_j}-t_Q^{\beta_j}}{\beta_j}+\cdots.
\end{aligned}
\]
The quadratic gauge does not change the lowest terms. By \cite[Theorem~2, Theorem~4]{NakayashikiYori2012} with
$\lambda_1=g$,
\[
\begin{aligned}
\left.\partial_{T_1}^{N}s_\lambda\right|_{T(A(P))}
&=\frac{N!\Delta_\beta}{\prod_{j\ge2}\beta_j!}\,t^g+\cdots,\\
\left.\partial_{T_1}^{N'}s_\lambda\right|_{T(A(P))-T(A(Q))}
&=(-1)^{g-1}\frac{N'!\Delta_\beta}{\prod_{j\ge2}(\beta_j-1)!}\,
(t_Pt_Q)^{g-1}(t_P-t_Q)+\cdots.
\end{aligned}
\]
Since $\partial_g=-(\varepsilon\ell_g)^{-1}\partial_{T_1}$, the leading
coefficients of $\partial_g^N\sigma(A(P))$ and $\partial_g^{N'}\sigma(A(P)-A(Q))$ are $a_\infty$
and $a_{\mathrm{pair}}$. Those of $\tilde E(\infty,P)$ and $\tilde E(P,Q)$, meaning the coefficients
of $t^g$ and $(t_Pt_Q)^{g-1}(t_P-t_Q)$, are $1$
(\cref{prop:standard}; \cite[(61)]{NakayashikiYori2012}). Substituting into
\eqref{eq:det} gives the formula.
\end{proof}

\begin{remark}[Second-kind choices and quadratic gauges]\label{rem:gauge}
For $dr^H=dr+Hdu$, $H=H^\trans$, the coherent changes are
\begin{gather*}
 \sigma^H(u)=e^{-u^\trans Hu/2}\sigma(u),\qquad
 \wp^H_{ij}=\wp_{ij}+H_{ij},\\
 \tilde E^H(P,Q)=e^{-(A(P)-A(Q))^\trans H(A(P)-A(Q))/2}\tilde E(P,Q).
\end{gather*}
Likewise $\tilde E^H(\infty,P)=e^{-A(P)^\trans HA(P)/2}\tilde E(\infty,P)$.
The residue constants in \eqref{eq:rows} change with $dr$ and compensate
these Kleinian shifts, leaving each remainder unchanged. The sigma-only
identity of \cref{cor:closed} is also unchanged: lower derivatives vanish
on the two shift loci, while the remaining quadratic factors cancel by
\[
 MM^\trans+\sum_i\mu_i\mu_i^\trans+
 \sum_{i<j}(\mu_i-\mu_j)(\mu_i-\mu_j)^\trans
 =(K+1)\sum_i\mu_i\mu_i^\trans,\qquad M=\sum_i\mu_i.
\]
The linear terms in $z$ cancel because $M=\sum_i\mu_i$.
\end{remark}

\begin{remark}
We reserve \^Onishi's sharp and flat notation for the hyperelliptic mixed
derivatives \cite[\S6]{Onishi2005}:
\[
\sigma_\sharp=\sigma_{2,4,6,\ldots},\qquad
\sigma_\flat=\sigma_{3,5,7,\ldots},
\]
with indices at most $g$ and an empty list meaning $\sigma$ itself.
Nakayashiki--Yori's gap-sequence derivatives recover this rule
\cite[Example~3]{NakayashikiYori2012}; their repeated last-coordinate
derivatives used in \cref{cor:closed} are written explicitly.
On the curve and two-point difference loci, respectively, the two
choices are proportional. In genus two they coincide:
$\partial_2^N\sigma=\sigma_\sharp=\sigma_2$ and
$\partial_2^{N'}\sigma=\sigma_\flat=\sigma$.
In genus three $\sigma_\sharp=\sigma_2$, whereas $\partial_3^N\sigma=\sigma_{333}$;
their restrictions to the curve are proportional. For the two-point
conversion see \cite[Remark~3]{NakayashikiYori2012}.
\end{remark}

\begin{remark}[Removing additive row constants]\label{rem:constant-free}
For $k>g$ put $\gamma_{jk}=-\Res_\infty(\fr_j\varphi_k\nu)$ and
$\widehat\varphi_k=\varphi_k-\sum_{j=1}^g\gamma_{jk}\varphi_j$.
Since $\mathcal R_i(\varphi_j)=\delta_{ij}$, the rows for
$\widehat\varphi_k$ have no additive residue constants. These fixed
column operations preserve the determinant and leading pole orders.
They are specified relative to the chosen $dr$, as are the constants.
\end{remark}

\begin{corollary}[Point-derivative form]\label{cor:derivative-chains}
Let primes denote point differentiation defined by $df=f'\nu$;
equivalently, differentiation in a local parameter whose differential is
$\nu$. Thus
\[
x'=\frac{F_y}{c_\nu},\qquad y'=-\frac{F_x}{c_\nu},\qquad
f'=\frac1{c_\nu}\left(F_y\frac{\partial f}{\partial x}
-F_x\frac{\partial f}{\partial y}\right)
\]
for a polynomial $f(x,y)$. A column with entries $f^{(k)}(P_i)$
at the fixed top points has bottom entries
\begin{equation}\label{eq:trace-rows}
\mathcal R_j(f^{(k)})(z)
=\partial_j\sum_{a=1}^g
 f^{(k-1)}\bigl(Q_a(z)\bigr),
\qquad k\ge1,\quad 1\le j\le g,
\end{equation}
where $D(z)=Q_1(z)+\cdots+Q_g(z)$ and
$z=\sum_{a=1}^gA(Q_a(z))$. The $Q_a(z)$ move with $z$; the top points
$P_i$ stay fixed. Here $\partial_j=\partial/\partial z_j$, whereas each
prime is a point derivative evaluated at the indicated point.

The additive residue constants in \eqref{eq:rows} vanish for every
point-derivative column.
\end{corollary}

\begin{proof}
Put $R_j=\rem_z(f')_j$. Since $df=f'\nu$, the remainder definition at
$Q_a(z)$ gives $df=\sum_{j=1}^gR_jdu_j$. Summing along the moving points,
\[
d_z\sum_{a=1}^gf(Q_a(z))
=\sum_{j=1}^gR_j\sum_{a=1}^gdu_j(Q_a(z))
=\sum_{j=1}^gR_j\,dz_j,
\]
which proves \eqref{eq:trace-rows} for $k=1$; apply it to $f^{(k-1)}$
for general $k$. Finally,
$-\Res_\infty(\fr_j\,df)=\Res_\infty(f\,dr_j)=0$, because
$f\,dr_j$ has no poles away from $\infty$.
\end{proof}

\begin{lemma}[Complete derivative-chain columns]\label{lem:complete-chains}
A complete column list has one function at each nongap pole order,
ordered increasingly. Put $h=2g-1$. In each residue class modulo $h$,
choose the monomial $f_r$ of least positive pole order $v_r$.
Together with $1$, the chains $f_r,f_r',f_r'',\ldots$ give a complete list.
If $f_r\sim\ell(f_r)t^{-v_r}$, their leading coefficients are
\[
 \ell(f_r^{(k)})=\ell(f_r)\varepsilon^{-k}
           \prod_{i=0}^{k-1}(v_r+ih),\qquad k\geq0.
\]
The determinant formulas retain their form with these new leading
coefficients.
\end{lemma}
\begin{proof}
Since $\nu\sim-\varepsilon t^{2g-2}dt$, differentiation raises a positive
pole order by $h$ and multiplies its coefficient by its order divided by
$\varepsilon$. Since $h$ is the largest gap of $\langle n,s\rangle$,
the chain orders are exactly all positive nongaps. The ordered functions
are triangular combinations of the monomials, with the first $g$
columns unchanged. Their diagonal coefficients give precisely the change
in the product of leading coefficients.
\end{proof}

\section{Application to hyperelliptic BA mode powers}\label{sec:mode-powers}

We call a BA mode \emph{elementary} when its sigma shift is the Abel lift
of a single curve point; a composite shift is a sum of such lifts.
For odd-degree hyperelliptic curves, the same determinants can be expressed
in elementary BA mode powers, meaning products of opposite-shift partners,
and their first derivatives. This gives a
direct connection to the modal variables of the canonical systems in
\cite{HeskethBA}.

\begin{proposition}[Mode-power form]\label{prop:mode-powers}
Let $y^2=f(x)=4x^{2g+1}+\cdots$ be smooth, with
$du_j=x^{j-1}dx/y$, and use the Baker second-kind and sigma conventions of
\cref{def:sigma}, as in \cite{BEL1997}.
Write the Jacobi-inversion polynomial as
\[
P(z,x)=\prod_{a=1}^g(x-x(Q_a(z)))
=x^g-\sum_{j=1}^g\wp_{gj}(z)x^{j-1}.
\]
Choose $g$ fixed node points $E_k=(a_k,b_k)$ with distinct abscissae,
and put $\mu_k^E=A(E_k)$. These interpolation nodes need not coincide
with the determinant's top points $P_i$. The elementary BA pairs are
\[
\xi_k=c_k\frac{\sigma(z-\mu_k^E)}{\sigma(z)}e^{\boldsymbol r_k^\trans z},\qquad
\bar\xi_k=d_k\frac{\sigma(z+\mu_k^E)}{\sigma(z)}e^{-\boldsymbol r_k^\trans z},
\]
with fixed nonzero $c_k,d_k$. The bar labels the opposite-shift partner.
The hyperelliptic addition identity \cite{BEL1997,HeskethBA} gives
\[
 I_k=\xi_k\bar\xi_k=\kappa_kP(z,a_k),\qquad
 \kappa_k=c_kd_k\sigma_\sharp(\mu_k^E)^2,\qquad I_k'=\partial_{z_g}I_k.
\]
The exponential rates cancel. The prime in this section denotes the
distinguished Schr\"odinger direction $\partial_{z_g}$, not the point
derivative of \cref{cor:derivative-chains}. A physical propagation variable
may differ by a fixed scale, as in \cite{HeskethBA}.
Every $z$-dependent bottom-block entry
$\mathcal R_j(\varphi_{g+r})(z)$ in the determinant of
\cref{thm:main}, $1\le j\le g$, $1\le r\le K$, is polynomial in
$I_k,I_k'$, $1\le k\le g$, with coefficients depending only on fixed curve and node
data. No higher mode-power derivatives are required.
\end{proposition}

\begin{proof}
The values $P(z,a_k)=I_k/\kappa_k$ determine the monic degree-$g$
polynomial $P$ by interpolation, and $I_k'/\kappa_k$ determine
$P_{z_g}$, of degree at most $g-1$. Abel inversion gives
$\partial_{z_g}x(Q_a)=y(Q_a)/P_x(z,x(Q_a))$. Differentiating
$P(z,x(Q_a(z)))=0$ therefore gives
$P_{z_g}(z,x(Q_a))=-y(Q_a)$.
Consequently, the remainder polynomial of any column $\varphi(x,y)$ is
\[
R_\varphi(z,x)
=\operatorname{rem}_{P(z,x)}
 \bigl[\varphi(x,-P_{z_g}(z,x))\bigr]
=\sum_{j=1}^g\mathcal R_j(\varphi)(z)x^{j-1}.
\]
Interpolation has only fixed node denominators, and division by the monic
$P$ is polynomial in its coefficients. This proves the assertion for
every column and every determinant size.
\end{proof}

Multiplying the bottom $g$ rows by the node Vandermonde matrix
$[a_k^{j-1}]_{k,j=1}^g$ replaces their entries by
$R_\varphi(z,a_k)$ and multiplies the determinant by
$\Delta(a)=\prod_{k<l}(a_l-a_k)$. In particular, the columns $x^g$ and
$y$ become $a_k^g-I_k/\kappa_k$ and $-I_k'/\kappa_k$:
they are linear in the powers and their derivatives. Later columns are
generally polynomial in the same data.

When $K\geq g$, the nodes may be selected from the top points; this is
the natural choice when using powers of the same retained BA modes.
For $E_k=P_k$, subtracting the corresponding transformed bottom row
from its top row gives $I_k/\kappa_k$ in the $x^g$ column and
$y_k+I_k'/\kappa_k$ in the $y$ column. The supplementary material
\cite{Supplementary} develops the resulting compact power determinants.

\section{Examples}\label{sec:examples}

\subsection{Genus one}

In genus one the Abel map identifies the curve with its Jacobian, so the
free-vector distinction begins at $g\geq2$. For $y^2=4x^3-g_2x-g_3$,
$du=dx/y$, one has $x=\wp(u)$ and $x'=\wp'(u)$. The complete chain is
$1,x,x',x'',\ldots$, with nonconstant leading coefficients
$1!,2!,3!,\ldots$. The bottom row is their evaluation at $u=z$, giving
the classical Frobenius--Stickelberger determinant and its factorial
factor $\prod_{r=1}^K r!$. In particular $K=1$ reads
\[
 -\det\begin{pmatrix}1&\wp(\mu)\\1&\wp(z)\end{pmatrix}
 =\frac{\sigma(z+\mu)\sigma(z-\mu)}{\sigma(z)^2\sigma(\mu)^2}
 =\wp(\mu)-\wp(z).
\]

\subsection{Genus two}\label{sec:genus-two}

Take $y^2=f(x)=4x^5+\lambda_4x^4+\lambda_3x^3+\lambda_2x^2+\lambda_1x+\lambda_0$,
so $(n,s)=(2,5)$, $g=2$ and $c=4$. Take $c_\nu=2$, so that $\nu=dx/y$,
$du_1=dx/y$ and $du_2=x\,dx/y$, with
$dr_j=N_j(x)\,dx/(4y)$, where $N_1(x)=\lambda_3x+2\lambda_4x^2+12x^3$ and
$N_2(x)=4x^2$. The monomials are $1,x,x^2,y,x^3,xy,x^4,\dots$, with $\ell_y=2$
and all other $\ell_k=1$; $x=t^{-2}$ at $\infty$. Here $\beta=(3,1)$,
$\lambda=(2,1)$, $N=1$, $N'=0$ and $\varepsilon=\ell_2=1$. The Baker
normalization gives $c_\sigma=1$, $\sigma_\sharp=\sigma_2$, $\sigma_\flat=\sigma$ and
$a_\infty=a_{\mathrm{pair}}=-1$:
$\tilde E(\infty,P)=-\sigma_2(A(P))$ and $\tilde E(P,Q)=\sigma(A(Q)-A(P))$.
\Cref{thm:main}(i) gives the rows in order of pole order:
\begin{align*}
\mathcal R_j(x^2)(z)&=\wp_{2j}(z),\qquad \mathcal R_j(y)(z)=\wp_{22j}(z),\\
\mathcal R_j(x^3)(z)&=\tfrac16\wp_{222j}(z)+\tfrac13\wp_{1j}(z)-\tfrac{\lambda_4}6\wp_{2j}(z)-\tfrac{\lambda_3}{12}\delta_{2j},\\
\mathcal R_j(xy)(z)&=\tfrac1{12}\wp_{2222j}(z)+\tfrac23\wp_{12j}(z)-\tfrac{\lambda_4}{12}\wp_{22j}(z),\\
\mathcal R_j(x^4)(z)&=\tfrac1{120}\wp_{22222j}(z)+\tfrac16\wp_{122j}(z)-\tfrac{\lambda_4}{24}\wp_{222j}(z)
-\tfrac{\lambda_4}{15}\wp_{1j}(z)\\
&\quad+\tfrac{2\lambda_4^2-9\lambda_3}{60}\wp_{2j}(z)
-\tfrac{\lambda_1}{20}\delta_{1j}+\tfrac{\lambda_3\lambda_4-6\lambda_2}{60}\delta_{2j}.
\end{align*}
The constants are the residues $-\Res_\infty(\fr_j\,\varphi\,dx/y)$. The
$\lambda$-free parts are Schur polynomials: the rows of $x^3$, $xy$ and $x^4$ are
$-\partial_jp_3$, $-2\partial_jp_4$ and $-\partial_jp_5$ applied to $\log\sigma$,
where $p_k=p_k(\partial_2,0,\tfrac13\partial_1,0,0,\dots)$ and the factor $2$ is
$\ell_{xy}$.

\begin{example}[Three points in genus two]\label{ex:genus2}
For curve points $P_i=(x_i,y_i)$, $\mu_i=A(P_i)$ ($i=1,2,3$) and $z\in\C^2$,
\[
\det\begin{pmatrix}
1&x_1&x_1^2&y_1&x_1^3\\
1&x_2&x_2^2&y_2&x_2^3\\
1&x_3&x_3^2&y_3&x_3^3\\
1&0&\wp_{12}(z)&\wp_{122}(z)&\tfrac16\wp_{1222}(z)+\tfrac13\wp_{11}(z)-\tfrac{\lambda_4}6\wp_{12}(z)\\[2pt]
0&1&\wp_{22}(z)&\wp_{222}(z)&\tfrac16\wp_{2222}(z)+\tfrac13\wp_{12}(z)-\tfrac{\lambda_4}6\wp_{22}(z)-\tfrac{\lambda_3}{12}
\end{pmatrix}
\]
\[
=2\,\frac{\sigma(z+\mu_1+\mu_2+\mu_3)\prod_{i=1}^3\sigma(z-\mu_i)}{\sigma(z)^4}\,
\frac{\prod_{1\le i<j\le3}\sigma(\mu_j-\mu_i)}{\prod_{i=1}^3\sigma_2(\mu_i)^4},
\]
by \cref{cor:closed} with $g=2$, $K=3$, $(-1)^{gK}=1$,
$\prod_{k=3}^5\ell_k=\ell_y=2$ and $a_\infty^{12}a_{\mathrm{pair}}^{-3}=-1$. Reversing the three pair
differences absorbs this sign because $\sigma$ is odd.
\end{example}

\emph{Derivative-chain form.} Here $x'=y$ and $y'=f_x(x)/2$,
where $f_x=df/dx$ and primes denote point differentiation. In genus two, Jacobi
inversion gives
\[
\sum_{a=1}^2x(Q_a(z))=\wp_{22},\qquad
\sum_{a=1}^2y(Q_a(z))=2\wp_{122}+\wp_{22}\wp_{222},\qquad
\sum_{a=1}^2x(Q_a(z))^2=\wp_{22}^2+2\wp_{12},
\]
with all Kleinian functions at $z$.

The correspondence of \cref{cor:derivative-chains} is
\[
\begin{array}{c|l}
\text{top entry at }P_i&\text{entry in bottom row }j\\ \hline
x'(P_i)&\partial_j\wp_{22}(z)\\[3pt]
x''(P_i)&\partial_j\!\left(\wp_{22}(z)\partial_2\wp_{22}(z)
                       +2\partial_2\wp_{12}(z)\right)\\[3pt]
(x^2)'(P_i)&\partial_j\!\left(\wp_{22}(z)^2+2\wp_{12}(z)\right)
\end{array}
\]
with $j=1,2$. These polynomial expressions also have the linear
higher-derivative forms supplied by \cref{thm:main}(i).
The three chains start with $x,x^2,x^3$. After constant rescalings, their ordered
columns begin
\[
1,\ x,\ x^2,\ x',\ x^3,\ \tfrac12(x^2)',\ \tfrac1{10}x'',\
\tfrac13(x^3)',\dots .
\]
For instance $\tfrac12(x^2)'=xy$, while
\[
\tfrac1{10}x''=x^4+\tfrac{\lambda_4}{5}x^3
+\tfrac{3\lambda_3}{20}x^2+\tfrac{\lambda_2}{10}x+\tfrac{\lambda_1}{20},
\qquad
\mathcal R_j(\tfrac1{10}x'')
=\tfrac1{120}\partial_2^4\wp_{2j}+\tfrac16\partial_2^2\wp_{1j}
-\tfrac{\lambda_4}{120}\partial_2^2\wp_{2j}.
\]
Earlier-column additions can further clean the displayed determinant.
Put $\Phi(x)=x^3+\lambda_4x^2/6+\lambda_3x/12$.
The $\lambda_3x/12$ correction removes the additive residue constant as in
\cref{rem:constant-free}; the $\lambda_4x^2/6$ correction also removes a
curve-coefficient multiple of a Kleinian term.
Adding $\lambda_4/6$ times column 3 and $\lambda_3/12$ times column 2
to column 5 of \cref{ex:genus2} gives the equivalent left-hand side
\[
\det\begin{pmatrix}
1&x_1&x_1^2&x_1'&\Phi(x_1)\\
1&x_2&x_2^2&x_2'&\Phi(x_2)\\
1&x_3&x_3^2&x_3'&\Phi(x_3)\\
1&0&\wp_{12}&\wp_{122}&\tfrac16\wp_{1222}+\tfrac13\wp_{11}\\
0&1&\wp_{22}&\wp_{222}&\tfrac16\wp_{2222}+\tfrac13\wp_{12}
\end{pmatrix},
\]
with the same sigma-product right-hand side. All Kleinian entries in
this paragraph are at $z$.

\emph{Mode-power form.} For example, let $\mathcal S(z)$ denote the sigma-product right-hand side
of \cref{ex:genus2}. Put $J_k=I_k/\kappa_k$,
$J_k'=I_k'/\kappa_k$ and
$q=a_1+a_2-(J_2-J_1)/(a_2-a_1)=\wp_{22}(z)$. Then that example becomes
\[
\det\begin{pmatrix}
1&x_1&x_1^2&y_1&x_1^3\\
1&x_2&x_2^2&y_2&x_2^3\\
1&x_3&x_3^2&y_3&x_3^3\\
1&a_1&a_1^2-J_1&-J_1'&a_1^3-(a_1+q)J_1\\
1&a_2&a_2^2-J_2&-J_2'&a_2^3-(a_2+q)J_2
\end{pmatrix}
=(a_2-a_1)\mathcal S(z).
\]
Thus the total-shift BA product has a determinant representation whose
only moving data are elementary mode powers and their first derivatives.

\subsection{A trigonal example}

Let $X$ be the $(3,4)$ curve $y^3=x^4+\lambda_9x+\lambda_{12}$, smooth of
genus three when $x^4+\lambda_9x+\lambda_{12}$ has distinct roots. Here
$\nu=dx/(3y^2)$, $du=(1,x,y)\,\nu$, so that $u_1,u_2,u_3$ have weights
$5,2,1$, and $dr=(5x^2y,\ 2xy,\ x^2)\,\nu$, which satisfies
\eqref{eq:legendre}. Among polynomial weight-homogeneous choices there
is no holomorphic correction: no monomial in $\lambda_9,\lambda_{12}$,
of weights $-9,-12$, has weight $-(\beta_i+\beta_j)$. Thus this basis gives the
Schur sigma convention of \cite{NakayashikiYori2012}. At $\infty$, $x=t^{-3}$ and $y=t^{-4}(1+O(t))$.
For $j=1,2,3$, \cref{thm:main}(i) gives
\begin{align*}
\mathcal R_j(x^2)(z)&=\wp_{3j}(z),\\
\mathcal R_j(xy)(z)&=\tfrac12\wp_{33j}(z)+\tfrac12\wp_{2j}(z),\\
\mathcal R_j(y^2)(z)&=\tfrac16\wp_{333j}(z)+\tfrac12\wp_{23j}(z),\\
\mathcal R_j(x^3)(z)&=\tfrac1{24}\wp_{3333j}(z)+\tfrac14\wp_{233j}(z)+\tfrac18\wp_{22j}(z)-\tfrac{\lambda_9}4\,\delta_{1j}.
\end{align*}
For $k=1,2,3,4$ these rows are
\[
-\partial_j\,p_k\bigl(\partial_3,\tfrac12\partial_2,0,0,\tfrac15\partial_1\bigr)\log\sigma ,
\]
plus the single constant $-\Res_\infty(\fr_1\,x^3\nu)=-\lambda_9/4$.

Here $\beta=(5,2,1)$, $\lambda=(3,1,1)$, $N=2$, $N'=0$ and
$\varepsilon=\ell_3=c_\sigma=1$. The derivatives are $\partial_3^N\sigma=\sigma_{33}$,
$\partial_3^{N'}\sigma=\sigma$ and $a_\infty=a_{\mathrm{pair}}=1$:
\[
\tilde E(P,Q)=\sigma(A(P)-A(Q)),\qquad
\tilde E(\infty,P)=\sigma_{33}(A(P))=\sigma_2(A(P)).
\]
For the last equality, \cite[Theorem~10]{NakayashikiYori2012} gives
the curve-point derivative $\sigma_2$. On $A(X)$, $\sigma_2$ has the same
multipliers and zeros as $\sigma_{33}$, since $\sigma$ and $\sigma_3$ vanish
there, and the same leading term $t^3$.

\begin{example}[Four points on a trigonal curve]\label{ex:trigonal}
For curve points $P_i=(x_i,y_i)$, $\mu_i=A(P_i)$ ($i=1,\dots,4$), and $z\in\C^3$, with the rows $\mathcal R_j(\varphi)(z)$ above,
\[
\det\begin{pmatrix}
1&x_1&y_1&x_1^2&x_1y_1&y_1^2&x_1^3\\
\vdots&\vdots&\vdots&\vdots&\vdots&\vdots&\vdots\\
1&x_4&y_4&x_4^2&x_4y_4&y_4^2&x_4^3\\
1&0&0&\mathcal R_1(x^2)(z)&\mathcal R_1(xy)(z)&\mathcal R_1(y^2)(z)&\mathcal R_1(x^3)(z)\\
0&1&0&\mathcal R_2(x^2)(z)&\mathcal R_2(xy)(z)&\mathcal R_2(y^2)(z)&\mathcal R_2(x^3)(z)\\
0&0&1&\mathcal R_3(x^2)(z)&\mathcal R_3(xy)(z)&\mathcal R_3(y^2)(z)&\mathcal R_3(x^3)(z)
\end{pmatrix}
\]
\[
=\frac{\sigma\bigl(z+\sum_{i=1}^4\mu_i\bigr)\prod_{i=1}^4\sigma(z-\mu_i)}{\sigma(z)^5}\,
\frac{\prod_{1\le i<j\le4}\sigma(\mu_i-\mu_j)}{\prod_{i=1}^4\sigma_2(\mu_i)^5},
\]
by \cref{cor:closed} with $g=3$, $K=4$, $(-1)^{gK}=1$, all $\ell_k=1$ and
$a_\infty=a_{\mathrm{pair}}=1$.
\end{example}

\emph{Derivative-chain form.} With $\nu=dx/F_y$ the point derivative
satisfies $x'=3y^2$ and $y'=4x^3+\lambda_9$. The five seed monomials are
$x,y,x^2,xy,x^2y$. After constant rescalings, the first seven columns are
\[
1,\ x_i,\ y_i,\ x_i^2,\ x_iy_i,\ \tfrac13x_i',\ \tfrac14y_i'.
\]
The sixth column is unchanged, and the seventh is $x_i^3+\lambda_9/4$.
Their bottom entries, for $j=1,2,3$, are
\[
\begin{array}{c|l}
\text{top entry}&\text{entry in bottom row }j\\ \hline
x_i'/3&\tfrac16\partial_3^2\wp_{3j}(z)+\tfrac12\partial_3\wp_{2j}(z)\\[3pt]
y_i'/4&\tfrac1{24}\partial_3^3\wp_{3j}(z)+\tfrac14\partial_3^2\wp_{2j}(z)
+\tfrac18\wp_{22j}(z)
\end{array}
\]
Consequently \cref{ex:trigonal} has the same right-hand side with these
top columns and the constant-free seventh column downstairs. This is
the column operation $C_7\mapsto C_7+(\lambda_9/4)C_1$.

\section{Discussion}\label{sec:discussion}

The identities developed here retain the curve-point rows characteristic
of \^Onishi's higher-genus determinant formulas \cite{Onishi2005}, while
allowing one argument vector to vary freely in $\C^g$ rather than remain
an Abel lift of a curve point. The explicit Kleinian rows make this separation
usable: for a specified curve and determinant size, they are obtained by
finite series expansions and residue calculations. This construction is
suited to symbolic implementation, returning both the sigma product and
its determinant expression without solving Jacobi inversion.

A principal application is the study of products of BA modes. The sigma
side contains the elementary factors and their total-shift composite,
while the determinant expresses their product through Abelian functions
of the evolving argument. Such expressions allow mixing terms to be
compared with modal-power terms in a Hamiltonian, and can expose the
cancellations underlying conservation laws. For odd-degree hyperelliptic
curves, \cref{prop:mode-powers} makes this connection explicit: elementary
mode powers at $g$ distinct nodes and their first derivatives determine
every bottom-row entry, for any determinant size. Polynomial division then
provides an algebraic route between the Kleinian and modal descriptions.

For other $(n,s)$ curves, the Kleinian determinant remains available, as
the trigonal example illustrates. An important next question is which
mode powers and derivatives suffice to reconstruct its rows with comparable
simplicity. The supplementary material gives sufficient trigonal
first-derivative frames and a smaller genus-three separation criterion. The separated form also provides a starting point for studying
even and odd combinations of mixing products and their role in constructing
closed coupled-mode equations.

The accompanying supplementary material \cite{Supplementary} develops
extensions to several unrestricted argument vectors and several pole
points on general curves. It also records further hyperelliptic and
trigonal power-coordinate identities, rational single-mixing equations,
and their supporting symbolic and numerical checks.

\section*{Code availability}

The source code, parameters and recorded results for the symbolic and
numerical checks are available at
\url{https://github.com/HeskethGD/gen-det-ids}, together with the
manuscript and supplementary material. The checks use Generapy~0.3.0
\cite{Generapy}; the repository records the dependency versions and
instructions for reproducing them.

\section*{AI assistance and validation}

The author conceived the research project, formulated its mathematical
objectives and directed its development. OpenAI and Anthropic AI systems
assisted with mathematical exploration and proof development, manuscript
preparation, and symbolic and numerical verification code, and provided
iterative manuscript reviews. The author selected the material presented
and takes responsibility for the final manuscript.

The work was checked through iterative mathematical review, exact
symbolic calculations of selected coefficients and residue constants,
and numerical comparisons of determinant expressions with direct
Generapy sigma-function evaluations and independent divisor
interpolation. Hyperelliptic and trigonal examples were tested; the
verification code and recorded results distinguish absolute-constant
checks from prefactor-constancy checks.

Generapy was also developed by the author with AI assistance. Its
automated validation includes unit tests against known mathematical
identities, integration tests connecting its curve, Abelian-integral,
theta-function and Kleinian-function components, and numerical
reproductions of results from the published literature. The tests and
literature-based examples are available in the Generapy repository and
documentation \cite{Generapy}.

\end{document}